\documentclass[12pt]{article}
\usepackage{amsmath,epsfig,graphicx,amssymb}

\newcommand{\beq}{\begin{equation}}
\newcommand{\eeq}{\end{equation}}
\newcommand{\ben}{\begin{eqnarray}}
\newcommand{\een}{\end{eqnarray}}
\newcommand{\bes}{\begin{subequations}}
\newcommand{\ees}{\end{subequations}}
\newcommand{\bFig}{\begin{figure}}
\newcommand{\eFig}{\end{figure}}

\date{}
\begin{document}

\title{Nonquantum Cognition}
\author{Partha Ghose\footnote{partha.ghose@gmail.com} \\
Centre for Astroparticle Physics and Space Science (CAPPS),\\Bose Institute, \\ Block EN, Sector V, Salt Lake, Kolkata 700 091, India}
\maketitle
\begin{abstract}
The Hilbert space structure of classical field theory is proposed as a general theoretical framework to model human cognitive processes which do not often follow classical (Bayesian) probability principles. This leads to an extension of the circumplex model of affect and a Poincar\'{e} sphere representation. A specific toy field theoretic model of the brain as a coherent structure in the presence of noise is also proposed that agrees qualitatively with Pavlovian fear conditioning studies.
\end{abstract}
\section{Introduction}
Quantum cognition is an emergent area of cognitive modeling that is novel and has many advantages over theories based on classical logic and classical probability theory \cite{aerts, pothos}. The main motivation for searching beyond the limits of classical logic and classical probability theory as a basis for modeling cognitive processes is empirical evidence such as the `Guppy effect' \cite{osh}. The overextension and underextension of membership weights of items that it implies cannot be explained by any classical model \cite{tversky,hampton,gabora}. The central aspect of quantum mechanics that the proponents of `quantum cognition' exploit to overcome the inadequacies of classical theory is essentially its Hilbert space structure that naturally incorporates superposition, interference, incompatibility, contextuality and order, and entanglement. The claim is that this structure lies at the origin of specific effects in cognition related to the way in which concepts and their combinations carry and influence thir meaning. The example given by Osheron and Smith is the pair of concepts {\em Pet} and {\em Fish} and their conjunction {\em Pet-and-Fish} or their disjunction {\em Pet-or-Fish}. They observed that while an exemplar or item such as Guppy was a very typical example of Pet-Fish, it was neither a very typical example of Pet nor of Fish. This psychological behaviour is unexpected in classical theory. What is shown is that quantum modeling gives a more natural and satisfactory account of such psychological/cognitive processes, and reduces to classical modeling in special circumstances \cite{sozzo}. Two other examples where classical modeling is found to be inadequate are: (1) a person who is neither happy nor unhappy can be in a superposition of happy and unhappy states which is not permitted in classical modeling; (2) an empirical answer to two incompatible questions such as whether Linda is both a feminist and a bank teller has been found to be inconsistent with the predictions of classical probability theory.

Although no explicit use has been made of any quantum physics as such in the `quantum cognition' approach, only of its abstract mathematical structure, the fundamental question that has not been addressed is: how does this typically quantum structure survive in macroscopic systems like the brain which are thermodynamically open? Quantum systems are predicted to be fragile and subject to rapid decoherence that quenches their quantum properties \cite{zurek}. In this paper I wish to point out that there is no need to use the word `quantum' at all in invoking the Hilbert space structure to model cognitive processes because paradigmatic classical theories such as Maxwellian electrodynamics also have this structure \cite{ghose1}. As is well known, the superposition principle holds in electrodynamics and is the basis of all interference phenomena such as in Young's double-slit experiment. In all such phenomena the phase of the wave plays a crucial role and makes the wave function complex valued. Even nonseparability, which is the hallmark of quantum entanglement, occurs in classical optics and leads to violation of Bell-like inequalities \cite{spreeuw,simon,holleczek,borges,gabriel}. It has been shown that these novel states of classical light violate {\em noncontextuality} which is a straightforward generalization of an apparently innocuous realism about measurement, namely the conventional classical notion that a physical reality must be independent of how it is measured. Put more technically, noncontextuality is the requirement that the result of a measurement is predetermined and not affected by how the value is measured, i.e. not affected by previous or simultaneous measurement of any other compatible or co-measureable observable. The path/position of a classical light beam and its polarization turn out to be {\em contextual} variables for these novel states \cite{ghose2}.

There is a mistaken belief among many persons that Hilbert space structure is specific to quantum mechanics and distinguishes it from classical physics. This is far from the truth. In mathematics Hilbert space refers to any linear vector space with a complex valued inner product that allows length and angle to be defined.  Hilbert spaces include the familiar Euclidean space, spaces of square-integrable functions, spaces of sequences, Sobolev spaces consisting of generalized functions, and Hardy spaces of holomorphic functions. Hilbert spaces can be used to study the harmonics of vibrating strings. There is nothing quantum about these spaces.

In the following sections I will first give a brief introduction to the essential features of a Hilbert space (Section 2), clarify the specific role of Hilbert spaces in quantum mechanics (Section 3), and then show in what sense classical field theories have a Hilbert space structure (Section 4). In Secion 5 I propose a heuristic toy model of the brain that incorporates the possibility of coherence and entanglement in the presence of noise with a view to developing a scenario to confront the basic hypothesis (of the Hilbert space structure of cognitive processes) to critical experimental tests. 

\section{Hilbert Space}
Let $\cal{H}$ be a complex linear vector space of dimension $n$ spanned by the complete set of basis vectors $\{\vert x_i\rangle\}$  ($i = 1,2,...,n$) and its dual $\cal{H}^*$ by the complete set of basis vectors $\{\langle y_j\vert\}$ so that every $\vert X\rangle \in \cal{H}$ can be expressed as $\vert X\rangle = \sum_{i=1}^n c_i \vert x_i\rangle$ and every $\langle Y\vert \in \cal{H}^*$ as $\langle Y\vert = \sum_{i=1}^n d_i \langle y_i \vert$ where $c_i$ and $d_i$ are complex coefficients.  The inner product on $\cal{H}$ must be specified so as to have the following properties:
\begin{enumerate}
\item
$\langle X\vert X\rangle\geq 0$\,\, $\forall \vert X\rangle \in\cal{H}$ and $\langle X\vert X\rangle = 0$ iff $\vert X\rangle = 0$;
\item
$\langle a Y\vert b X\rangle = a^* b \langle Y\vert X\rangle$\,\, $\forall \vert X\rangle \in \cal{H}$ and $\forall \langle Y\vert \in \cal{H}^*$; and
\item
$\langle Y\vert X\rangle  = \overline{\langle X\vert Y\rangle}$\,\, $\forall \vert X\rangle \in \cal{H}$ and $\forall \langle Y\vert \in \cal{H}^*$.
\end{enumerate}
The length of a vector $\vert X\rangle$ is usually denoted by $\vert \vert X \vert \vert$ and the inner product $\langle Y\vert X\rangle = \vert \vert X \vert \vert\, \vert \vert Y \vert \vert\, {\rm cos}\theta$ in the real case. Hence, $\vert\vert X\vert\vert = \sqrt{\langle X\vert X\rangle} = \sum_i \vert c_i\vert^2$. The basis vectors are usually chosen to be orthonormal, i.e. $\langle x_j\vert x_i\rangle = \delta_{ij}$. Being endowed with an inner product and an induced length or norm, $\cal{H}$ is called a Hilbert space. For example, the space $\mathbb{C}^n$ of n-tuples of complex numbers $z = (z_1, z_2, ...,z_n)$ and $z^\prime = (z_1^\prime, z_2^\prime, ...,z_n^\prime)$ is a Hilbert space under the inner product defined by $\langle z, z^\prime\rangle = \sum_{k=1}^n  z_k \bar{z}_k^\prime$. The square integrable functions on $L^2$ also form a Hilbert space. For any $f$ and $g$ in $L^2$ one defines the inner product $\langle f, g\rangle = \int_{\Omega}f(x) \overline{g(x)} dx$. Hence, $\langle f, f\rangle < \infty$. 

{\flushleft{{\em Superposition principle}}}

The linearity of Hilbert spaces allows vectors to be superposed to form other vectors in the same space. Thus, for $\vert X\rangle, \vert Y\rangle \in {\cal{H}}$, $\vert X\rangle + \vert Y\rangle \in {\cal{H}}$. An example is the interference of two coherent classical waves at a point $x$. Let $F_1(x, t) = a e^{i(\phi_1(x) - \omega t)}$ and $F_2(x, t) = b e^{i(\phi_2 (x) -\omega t)}$ be two waves at $x$ at time $t$. Then, the sum of the two waves is $G(x, t) = F_1(x, t) + F_2(x, t)$ and the intensity at $x$ is given by $I(x) = \langle G(x, t), G(x, t)\rangle = \int G(x, t) \overline{G(x, t)} dt = a^2 + b^2 + 2 ab\, {\rm cos} (\phi_1 - \phi_2)$. The second term gives rise to the familiar interference pattern. 

The direct sum of two Hilbert spaces ${\cal{H}}_1$ and ${\cal{H}}_2$ (${\cal{H}}_1 \cap {\cal{H}}_2 = \emptyset$) of dimensions $n_1$ and $n_2$ is the set ${\cal{H}}_1 \oplus {\cal{H}}_2$ of pairs of vectors ($\vert h_1\rangle, \vert h_2\rangle)$ in ${\cal{H}}_1$ and ${\cal{H}}_2$ with the operations 
\begin{enumerate}
\item
$(\vert h_1\rangle, \vert h_2\rangle) + (\vert h_1^\prime\rangle, \vert h_2^\prime\rangle) = (\vert h_1\rangle + \vert h_1^\prime\rangle, \vert h_2\rangle + \vert h_2^\prime\rangle)$
\item
$c(\vert h_1\rangle, \vert h_2\rangle) = (c\vert h_1\rangle, c\vert h_2\rangle)$ 
\end{enumerate}
Its dimension is ($n_1 + n_2$).

{\flushleft{{\em Projection operators}}}
  
One can define projection operators $\pi_i = \vert x_i\rangle\langle x_i\vert$ for every basis vector $\vert x_i\rangle$ which are hermitian ($\pi_i^\dagger = \pi_i$) and idempotent ($\pi_i. \pi_i = \pi_i$). They have the important property that $\pi_i (1 - \pi_i) = 0$, i.e. a projector and its complement are orthogonal. Thus, $\pi_i\vert X\rangle = \sum_j c_j \vert x_i\rangle\langle x_i\vert \vert x_j\rangle = c_i \vert x_i\rangle$, and hence $c_i = \langle x_i\vert\pi_i\vert X\rangle$ and $\vert c_i\vert^2 = \langle X\vert\pi_i\vert X\rangle$. Thus, $\vert c_i\vert$ is a measure of the membership of $\vert x_i\rangle$ in $\vert X\rangle$.

{\flushleft{{\em Linear transformations and choice of basis}}}

An important characteristic of vector spaces is the complete freedom to choose any set of basis vectors related by unitary transformations. Thus, for example, one can express $\vert X\rangle = \sum_i c_i \vert x_i\rangle$ or as $\vert X\rangle = \sum_i c_i^\prime \vert x_i^\prime\rangle$ provided $\vert x_i^\prime\rangle = \sum_j S_{ij} \vert x_j\rangle$ with $S$ a unitary matrix, i.e. $S^\dagger S = S S^\dagger = 1$ so that $S^\dagger = S^{-1}$. Operators $\cal{O}$ (represented by $n \times n$ matrices in the old basis) acting on $\cal{H}$ then transform to the new basis as ${\cal{O^\prime}} = S {\cal{O}}S^{-1}$ which is a similarity transformation.

{\flushleft{{\em Operators on Hilbert space and groups}}}

Bounded operators on a Hilbert space can form {\em non-Abelian} Lie groups and generate their algebras. A simple example is the rotation group $SO(3)$ in Euclidean space $\mathbb{R}^3$. As is well known, rotations along the three Cartesian axes do not commute, and this non-commutation has nothing to do with quantum mechanical incompatibility and uncertainty. Rather, it reflects certain symmetry properties of Euclidean space, and is correctly described by the non-Abelian property of the group $SO(3)$. $SO(3)$ has a universal covering group $SU(2)$, the group of all $2\times 2$ unitary matrices with complex elements and determinant $1$. Although the group $SU(2)$ is used in quantum physics to describe particles with spin $\frac{1}{2}\hbar$ (fermions), the group itself is independent of quantum mechanics, and can be used to describe any two-valued objects. These two examples should suffice to make the point clear that non-Abelian groups of operators act naturally on Hilbert spaces, and their non-commutative structure has nothing to do with quantum physics. This property of Hilbert spaces should be borne in mind in modeling human cognitive processes.

{\flushleft{{\em Tensor products}}} 

Apart from the direct sum ${\cal{H}}_1 \oplus {\cal{H}}_2$ of two Hilbert spaces, one can also form tensor products ${\cal{H}} = {\cal{H}}_1 \otimes {\cal{H}}_2$ whose dimension is $(n_1 \times n_2)$. In fact, there is a mathematical theorem which states that {\em every pair of vector spaces has a tensor product} \cite{sternberg}. Let $\{\vert e_i\rangle \} (1 \leq i \leq n_1)$ and $\{\vert f_j\rangle\} (1 \leq j \leq n_2)$ be the sets of basis vectors of ${\cal{H}}_1$ and ${\cal{H}}_2$ respectively. Then, the set $\{\vert e_i\rangle \otimes \vert f_j\rangle\} ( \leq i < n_1, 1 \leq j < n_2)$ forms the basis of ${\cal{H}}$. Here, $\{\vert e_i\rangle \otimes \vert f_j\rangle\}$ stands for the $(n_1 \times n_2)$ matrix whose $i$th row consists of the ordinary products $\vert e_i\rangle \vert f_j\rangle (1 \leq j \leq n_2)$. Hence it has $n_1$ rows and $n_2$ columns. A typical element of ${\cal{H}}_1 \otimes {\cal{H}}_2$ would be $\sum_{i,j}c_{ij} (\vert e_i\rangle \otimes \vert f_j\rangle)$. Since this tensor product space is also a Hilbert space, one can define an inner product on it by

$$\langle (\langle e_i\vert \otimes \langle f_i\vert)(\vert e_j\rangle \otimes \vert f_j\rangle) \rangle = \langle e_i\vert e_j\rangle \langle f_i\vert f_j\rangle$$ 

Now consider two $2$-dimensional Hilbert spaces ${\cal{H}}_A$ and ${\cal{H}}_B$ with basis vectors $\vert e_A\rangle = (\vert h\rangle_A, \vert v\rangle_A)$ and $\vert e_B\rangle = (\vert h\rangle_B, \vert v\rangle_B)$. One can construct the following four states from them:

\ben
\vert \Phi^+\rangle &=& \frac{1}{\sqrt{2}} [\vert h\rangle_A\otimes \vert h\rangle_B + \vert v\rangle_A\otimes \vert v\rangle_B ]\label{b1}\\
\vert \Phi^-\rangle &=& \frac{1}{\sqrt{2}} [\vert h\rangle_A\otimes \vert h\rangle_B - \vert v\rangle_A\otimes \vert v\rangle_B ]\label{b2}\\
\vert \Psi^+\rangle &=& \frac{1}{\sqrt{2}} [\vert h\rangle_A\otimes \vert v\rangle_B + \vert v\rangle_A\otimes \vert h\rangle_B ]\label{b3}\\
\vert \Psi^-\rangle &=& \frac{1}{\sqrt{2}} [\vert h\rangle_A\otimes \vert v\rangle_B - \vert v\rangle_A\otimes \vert h\rangle_B ]\label{b4}
\een
These states are of fundamental significance. When two states $\vert A\rangle = \sum_i c_i \vert e_i\rangle$ and $\vert B\rangle =\sum_j d_j \vert f_j\rangle$ are independent of one another, one can construct a product state $\vert X\rangle = \vert A\rangle \otimes \vert B\rangle = \sum_{i,j} c_i d_j \vert e_i\rangle \otimes \vert f_j\rangle$, and the state $\vert X\rangle$ is said to be `factorizable' into $\vert A\rangle$ and $\vert B\rangle$ which retain their identity, independence and separability. However, an inspection of (\ref{b1}, \ref{b2}, \ref{b3}, \ref{b4}) shows that the basis states $(\vert h\rangle_A, \vert v\rangle_A)$ and $(\vert h\rangle_B, \vert v\rangle_B)$ cannot be factored out of the four states $\vert \Phi^+\rangle, \vert \Phi^-\rangle, \vert \Psi^+\rangle, \vert \Psi^-\rangle$. Hence, none of these four states is factorizable. Such states may be called `entangled states' because the states from which they are constructed lose their identity, independence and separability in such states. The four states above form a special set of basis states which are maximally entangled, and may be called Bell-like states or the Bell-like basis. Notice that so far, no effort has been made to identify Hilbert spaces with any physical systems.

This ends our brief review of the salient mathematical features of Hilbert spaces that can be used to model unusual properties of systems, both classical and quantum, for which traditional scientific and logical methods based on Boolean algebra and classical probability theory are not adequate. 
 
\section{Hilbert Spaces in Quantum Mechanics}

In this section I will try to list the additional restrictions that are imposed on Hilbert spaces to get quantum mechanics which is so very different from Newtonian classical physics. 
\begin{enumerate}
\item
First, in order to have a probabilistic interpretation, quantum state vectors are all normalised (i.e. of unit norm). Hence, the set of all pure states corresponds to the unit sphere in Hilbert space, with the additional requirement that all vectors that differ only by a complex scalar factor (a phase factor) are identified with the same state. Thus, quantum mechanics operates on coset spaces and Grassmanian manifolds \cite{chatur}.

As a consequence, in quantum mechanics a physical state in general does not {\em possess} physical properties {\em before} measurement. For example, a single-photon state like $\vert X\rangle = c_1\vert H\rangle + c_2 \vert V\rangle$ with $\vert c_1\vert^2 + \vert c_2\vert^2 = 1$ cannot be said to possess any polarization $H$ or $V$. On measurement, however, the state is projected to either $\vert H\rangle$ with probability $\vert c_1\vert^2$ or $\vert V\rangle$ with probability $\vert c_2\vert^2$. This is not the case in classical electrodynamics which is deterministic and in which a state $\vert X\rangle = c_1\vert H\rangle + c_2 \vert V\rangle$ of light with classical amplitudes $c_1$ and $c_2$ {\em always} possesses a definite polarization given by the vector sum of $c_1\vert H\rangle$ and $c_2 \vert V\rangle$. $\vert c_1\vert$ and $\vert c_2\vert$ are here measures of the membership weights of $\vert H\rangle$ and $\vert V\rangle$ in the state $\vert X\rangle$. This {\em innocuous scientific realism} of the polarization states in classical electrodynamics, however, holds only for product states like $(c_1 \vert x\rangle + c_2 \vert y\rangle)\otimes \vert w\rangle$ but not for a superposition of product states like $c_1 \vert u\rangle \otimes \vert x\rangle + c_2 \vert v\rangle \otimes \vert y\rangle$. Birefringent crystals have been known since the 17th century to produce states of the form $c_1 u \otimes H + c_2 v \otimes V$ where $u$ is the path with $H$ polarization and $v$ is the path with $V$ polarization. This is indeed a polarization-path entangled state! I will return to this in more detail in a later section.
  
\item
Second, a linear and unitary equation of motion, the Schr\"{o}dinger equation, is postulated that specifies the time evolution of states in Hilbert space. 
\item
Third, all observables are represented by hermitian operators ${\cal{O}}^\dagger = {\cal{O}}$ on Hilbert space. Let $\vert \Psi\rangle = \sum_i c_i \vert \psi_i\rangle$ be a pure state, and let $\rho = \vert \Psi\rangle\langle \Psi\vert$ be the density operator which has the properties $\rho^2 = \rho$ and ${\rm Tr} \rho^2 = 1$. Then, the expectation value of ${\cal{O}}$ is given by $\bar{\cal{O}} = {\rm Tr} \rho \,{\cal{O}}$. The results of observations are obtained by measurements $M_i = \vert \psi_i\rangle\langle \psi_i\vert$ acting on the state $\vert \Psi\rangle: \hat{\rho} = \sum_i M_i \rho M_i^\dagger$. It is clear that ${\rm Tr}\hat{\rho}^2 < 1$. $\hat{\rho}$ is called the reduced density operator, and represents mixed states. It is straightforward to show that $\hat{\rho}_{ij} = \vert c_i\vert^2 \delta _{ij} $. This projection is additional to the linear and unitary time evolution. This is the measurement postulate. Its ad hoc and non-unitary character has spawned a plethora of interpretations of quantum mechanics \cite{int}.
\item
Fourth, every pair of canonical dynamical variables ($ p_i, q_j$) is postulated to be represented by hermitian operators ($\hat{p}_i, \hat{q}_j$) with the commutation rules $[\hat{p}_i, \hat{q}_j] = - i\hbar \delta_{ij}$ resulting in the famous Heisenberg uncertainty relations $\Delta q\, \Delta p \geq \hbar/2$. This is the canonical quantization postulate. These commutation relations vanish in the formal limit $\hbar \rightarrow 0$, and hence do not arise from a non-abelian character of the operators and are not intrinsic to Hilbert spaces.
\item
Fifth, entangled states in quantum mechanics are extremely {\em fragile}, i.e. the entanglement disappears very rapidly and the system decoheres and breaks into pieces when exposed to an environment \cite{zurek}. Such is not the case with classical electrodynamics in which coherence and entanglement/nonseparability are robust.

\end{enumerate}

We will now pass on to a brief description of the mathematical structure of classical field theories.

\section{Hilbert Space and Classical Field Theories}

In this section, I will repeat what I have written elsewhere by way of introducing the Hilbert space structure of classical electrodynamics. Classical electrodynamics is the paradigm of classical field theories in physics. That classical electrodynamics has a Hilbert space structure was first explicitly shown in 2001 \cite{ghose1}. Without getting involved in the details of that demonstration, let us simply note that two different Hilbert spaces are required for a complete description of an ordinary state in classical electrodynamics, namely a space $\hat{H}_{path}$ of square integrable functions that describe scalar optics and a two-dimensional space of polarization states $\hat{H}_{pol}$. These are disjoint Hilbert spaces, and hence a complete description of a state is given in terms of tensor products of states in these two Hilbert spaces: $\frac{1}{\sqrt{\vert A\vert^2}}\vert A\rangle\otimes \vert \lambda\rangle\in \hat{H}_{path}\otimes \hat{H}_{pol}$ where $A({\bf r},t) = \langle {\bf r},t\vert A\rangle$ are solutions of the scalar wave equation
\beq
\left[\nabla^2 - \frac{1}{c^2}\frac{\partial^2}{\partial t^2}\right]A({\bf r},t) = 0 
\eeq 
and $\vert \lambda\rangle$ is the vector \[\vert \lambda\rangle =\left(\begin{array}{c}
\lambda_1 \\\lambda_2
\end{array} \right) \] of the transverse polarizations $\lambda_1$ and $\lambda_2$. One can also write $\frac{1}{\sqrt{\vert A\vert^2}}\vert A\rangle\otimes \vert \lambda\rangle$ more conventionally as the Jones vector \[\vert J\rangle = \frac{1}{\sqrt{\langle J\vert J\rangle}}\left(\begin{array}{c}
 E_x\\ E_y
\end{array} \right) \] where $E_x = A_0 \hat{e}_x {\rm exp (i\phi_x)}$ and $E_y = A_0 \hat{e}_y {\rm exp (i\phi_y)}$ are the complex transverse electric fields, $\hat{e}_x$ and $\hat{e}_y$ are unit polarization vectors, and $\langle J\vert J\rangle = \vert E_x\vert^2 + \vert E_y\vert^2 =  A_0^2$ is the intensity $I_0$.
Given this mathematical structure of a tensor product Hilbert space, polarization-path entanglement is inevitable because, as we have seen, there is a mathematical theorem which states that {\em every pair of vector spaces has a tensor product}. The tensor product space is also a linear vector space. Hence, the existence of tensor product spaces resulting in entanglement is just as inevitable in classical optics as in quantum mechanics. It should be clear from this that classical electrodynamics is a lot more like quantum mechanics than is classical mechanics in which states are points in phase space. One can therefore construct the complete set of Bell-like states
\ben
\vert\Phi^+\rangle &=& \frac{A}{\sqrt{2I_0}}\left[\vert a\rangle\otimes\vert V\rangle+\vert b\rangle\otimes\vert H\rangle\right],\\
\vert\Phi^-\rangle &=& \frac{A}{\sqrt{2I_0}}\left[\vert a\rangle\otimes\vert V\rangle-\vert b\rangle\otimes\vert H\rangle\right],\\
\vert\Psi^+\rangle &=& \frac{A}{\sqrt{2I_0}}\left[\vert a\rangle\otimes\vert H\rangle+\vert b\rangle\otimes\vert V\rangle\right],\\
\vert\Psi^-\rangle &=& \frac{A}{\sqrt{2I_0}}\left[\vert a\rangle\otimes\vert H\rangle-\vert b\rangle\otimes\vert V\rangle\right],
\een
where $\vert a\rangle$ and $\vert b\rangle$ are non-intersecting paths that span a two-dimensional Hilbert space $H_{path} = \{\vert a\rangle, \vert b\rangle \}$.

Now, it is well known that classical fields are infinite collections of harmonic oscillators. This can be easily seen as follows. Consider the Hamiltonian of the electromagnetic field,
\beq
H = \frac{1}{8\pi}\int d^3 x  [{\bf E}^2 + {\bf B}^2].
\eeq
Identifying ${\bf E} \equiv 4\pi {\bf p}$ and ${\bf B} \equiv 4\pi {\bf q}$, this can be written as
\beq
H = \int d^3 x \frac{1}{2} [p^2 + q^2].
\eeq
One can discretise it and write it as
\beq
H = \frac{1}{2} \sum_a^{N=\infty}  [p_a^2 + q_a^2],\label{ham}
\eeq
which is an infinite collection of harmonic oscillators of different frequencies. 

The human brain is known to be a complex network of an incredibly large number (about tens of billions) of interconnected neural oscillators. It is therefore a system that should exhibit `collective oscillations' of various types depending on the boundary conditions. Such collective modes are called `normal modes', each of which is an independent harmonic oscillation of a characteristic frequency \cite{landau}. The most well known of these oscillations are the $\alpha, \beta, \theta, \delta$ brain waves recorded by electroencephalographs. There is also evidence of synchronous oscillations in the cerebral cortex \cite{col osc}. Such oscillations have been linked to cognitive states, such as awareness and consciousness \cite{engel, varela}. That clearly indicates that it should be possible to model the cognitive states of the human brain by the collective states of an infinite collection of harmonic oscillators, i.e. by a {\em classical field} with appropriate boundary conditions. Once this is realized, it becomes at once clear why human cognitive behaviour exhibits features of a Hilbert space structure, namely superposition of states, coherence, interference and even nonseparability/entanglement of its spatially separated regions responsible for coordinating `scattered mosaics of functionally specialized brain regions' \cite{varela}. 

It must be emphasized here that what I am proposing is a broad theoretical framework based on classical field theory and neurobiology to justify the use of modeling human cognition with apparently quantum probabilistic methods initiated by Aerts and others. It remains to work out more detailed dynamical models incorporating specific brain structures for every conceptual phenomena within this framework and confronting them with empirical data.  

\section{An Illustrative Example from Cognitive\\ Neurobiology}

The field theory that is relevant for modeling cognitive processes must be similar in its mathematical strucure to classical electrodynamics but obviously not identical to it. The complex processes that underlie and embody this structure are basically electrochemical and neurobiological in nature. Accordingly, let us use $N$ solutions $A_i (\{{\bf x}_i\},t)$ of the scalar field equations with appropriate boundary conditions to correspond to the $N$ brain regions with specialized functions like, for example, the amygdala, hippocampus, thalamus, prefrontal cortex, visual cortex, etc., and a $2$-dimensional Hilbert space to describe a pair of emotional states. These `internal' pairs of states will be described by a two-dimensional Hilbert space similar to the polarization Hilbert space in classical optics. In order to make probabilistic predictions for mental states, it is necessary to use only normalized states obtained by a simple division by the the total amplitude $\sqrt{\langle J\vert J\rangle}$ of the state. It must be borne in mind, however, that although the resulting probability theory will be effectively the same as quantum probability theory, it will not have the same fundamental significance. It will be similar in its foundations to the probability theory involved in certain deterministic hidden variable theories of quantum phenomena such as the de Broglie-Bohm theory in which the same statistical predictions as in quantum mechanics arise due to the lack of information about the hidden variables rather than to any fundamental indeterminacy in nature \cite{bohm}. In cognitive science, the complex neural networks and the correlates of mental states that underlie cognitive processes should be treated as {\em quasi-hidden} variables about which we do not have sufficient information. The phrase {\em quasi-hidden} is being used to emphasize the fact that unlike in the de Broglie-Bohm theory, these hidden variables are not really `hidden' and can be observed. There is in this case a complementarity between the field description and the neurobiological description, each having its own arena of validity and applicability that is exclusive and incompatible with the other. The field aspect may be regarded as an {\em emergent} phenomenon in a complex nonlinear system like the brain.
\begin{figure}
\centering
{\includegraphics[scale=0.4]{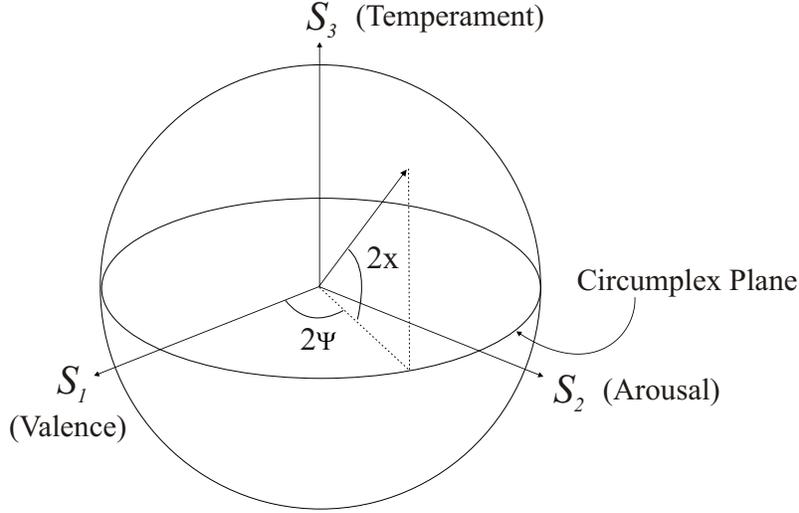}}
\caption{\label{Figure 1}{\footnotesize The Poincar\'{e} sphere for the extended circumplex model.}}
\end{figure}
\begin{figure}
\centering
{\includegraphics[scale=0.4]{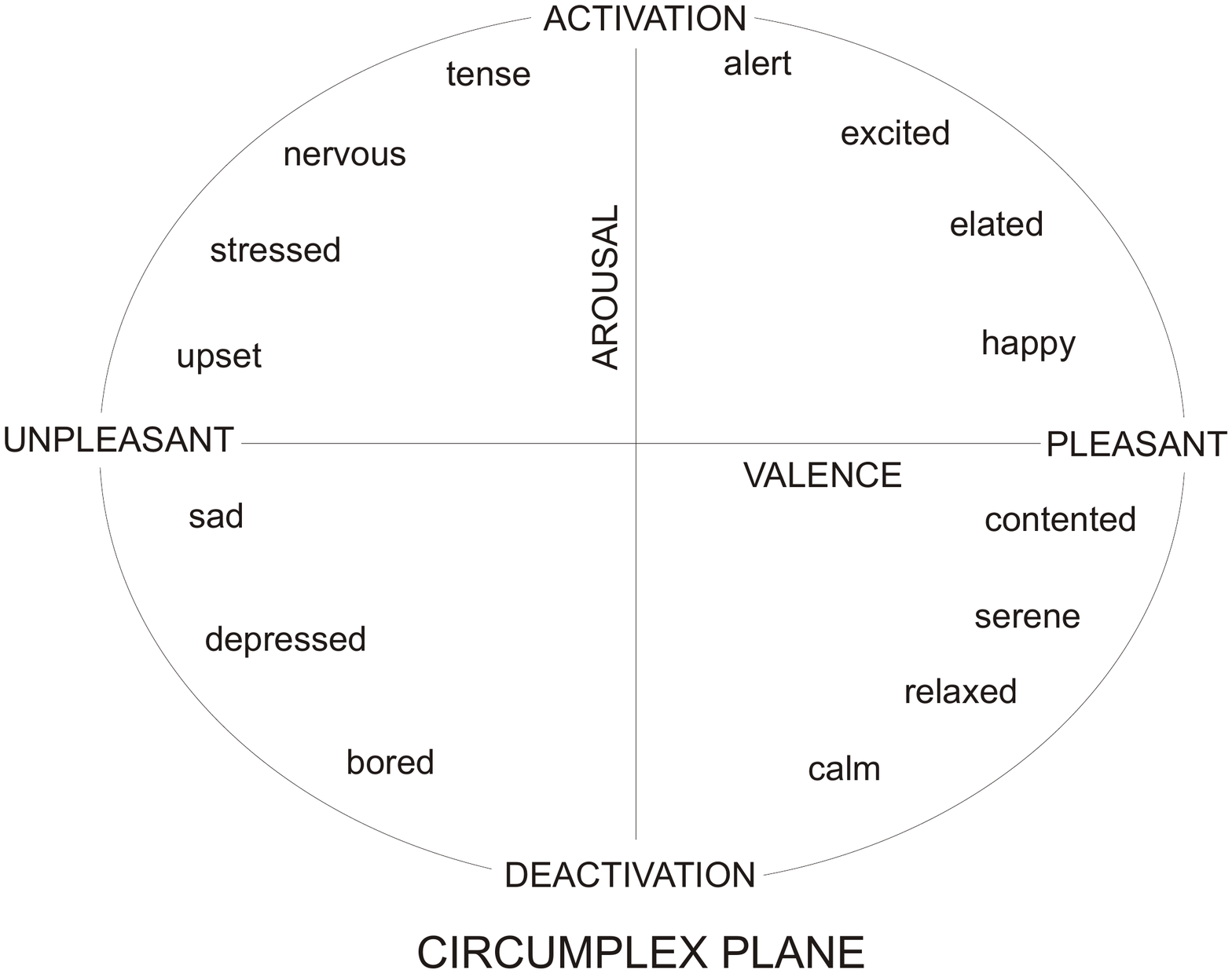}}
\caption{\label{Figure 2}{\footnotesize The circumplex plane.}}
\end{figure}

Modeling emotions by the mathematical structure of a polarization Hilbert space has the advantage of being able to use the Poincar\'{e} sphere for the specification of these states. Taking $X, Y, Z$ as the Cartesian axes, let us define the parameters
\ben
S_0^2 &=& S_1^2 + S_2^2 + S_3^2,\\
S_1 &=& S_0 \cos 2\chi \cos 2\psi,\\
S_2 &=& S_0 \cos 2\chi \sin 2\psi,\\
S_3 &=& S_0 \sin 2 \chi,
\een 
where $2\psi$ and $2\chi$ are the azimuth and polar angles in the Poincar\'{e} sphere \cite{bw} (Fig. 1). Then, $S_1$ is in the $X$ direction, $S_2$ in the $Y$ direction and $S_3$ in the $Z$ direction. Let the $X$-axis denote `valence', the $Y$-axis `arousal' and the $Z$-axis a third trait of affect which we tentatively take to be `temperament'.
This leads to an extended `circumplex model' of affect \cite{russell} in which, in addition to the two axes `valence' and `arousal', there is a third axis. {\em This is a necessary consequence of the assumed $2$-dimensional Hilbert space structure of affect}.  Thus, $S_1 > 0, S_2 > 0, S_3 > 0$ correspond to positive valence, positive arousal and positive temperament, and $S_1 < 0, S_2 < 0, S_3 < 0$ correspond to negative valence, negative arousal and negative temperament. Every point on this sphere can then represent a pure emotional state $\rho = \vert \psi \rangle \langle \psi\vert$, and every point inside the sphere a mixed emotional state described by the density matrix
\beq
\rho = \Sigma_i\, p_i \vert \psi_i \rangle \langle \psi_i\vert
\eeq
where $p_i$ are the probabilities with which the states $\vert \psi_i \rangle$ occur in a statistical ensemble of states. The expectation value or sample average of any observable ${\cal{O}}$ is given by
\beq
\langle {\cal{O}}\rangle = Tr (\rho {\cal{O}}) = \Sigma_i\, p_i \langle \psi_i\vert {\cal{O}}\vert \psi_i\rangle,
\eeq
i.e. the expectation value for the mixed state is the sum of the expectation values of ${\cal{O}}$ for each of the pure states $\vert \psi_i\rangle$, weighted by the probabilities $p_i$.

The standard circumplex model corresponds to the equatorial plane of the Poincar\'{e} sphere with the pure states sitting on the equator (Fig. 2). The antipodal points on the sphere correspond to mutually orthogonal states. 

One can define three matrices   
\beq
J_x = \left(\begin{array}{cc} 0 & \quad 1 \\ 1 & \,\,\,\,\,\,\,0\end{array}\right),\,\,\,\,J_y = \left(\begin{array}{cc} 0 & \quad -i \\ i & \,\,\,\,\,\,\,0\end{array}\right),\,\,\,\,J_z = \left(\begin{array}{cc} 1 & \quad 0 \\ 0 & \,\,\,\,\,-1\end{array}\right)
\eeq
which act on the emotional states and form the SU(2) Lie algebra $[J_i, J_j] = 2i \epsilon_{ijk} J_k$. They are identical with the Pauli matrices $\sigma_i$, but their interpretation is quite different from the standard interpretation of Pauli matrices as describing spin $\frac{1}{2}\hbar$ particles. These matrices represent operations on emotional states which are modelled to be like other two-valued entities in physics like intrinsic spin and polarization of classical light. The eigenstates of $J_x$ and $J_y$ are 
\ben
\vert x +\rangle &=& \frac{1}{\sqrt{2}}\left(\begin{array}{c} 1 \\ 1 \end{array}\right),\,\,\vert x -\rangle = \frac{1}{\sqrt{2}}\left(\begin{array}{c} 1 \\ -1 \end{array}\right),\\ \vert y +\rangle &=& \frac{1}{\sqrt{2}}\left(\begin{array}{c} 1 \\ i \end{array}\right),\,\,\vert y -\rangle = \frac{1}{\sqrt{2}}\left(\begin{array}{c} 1 \\ -i \end{array}\right),
\een
and they denote positive and negative states of `valence' and `arousal' respectively, while the eigenstates of $J_z$,
\beq
\vert z +\rangle = \left(\begin{array}{c} 1 \\ 0 \end{array}\right) {\rm and}\,\, \vert z -\rangle = \left(\begin{array}{c} 0 \\ 1 \end{array}\right)
\eeq
denote positive and negative states of `temperament'. 

There is some empirical evidence to support the need for an additional emotional state like temperament to augment the circumplex model, coming from the occurrence of comorbidity \cite{russell}. The `basic emotions' model \cite{ekman} is inappropriate as a starting point for our purpose, though it certainly has its own realm of validity within hard core neurobiology based on neuroimaging, and can be accommodated within our model.

We also assume that the brain is a very coherent structure but with some noise. Empirical evidence for coherence comes basically from the occurrence of psychophysiological homeostasis and studies recording coherent electrical signals in the brain and mapping neurobiological pathways and circuits \cite{mccraty}. With this background, we can write down a toy model of a typical brain state by
\ben
\rho_B (\{ A_i\}) &=& \Pi_{i=1}^N A_i (\{{\bf x}_i\},t) (\rho_W)_t\\
(\rho_W)_t &=& [ \gamma \rho_{ghz} + (1 - \gamma)\rho_{noise}]_t \label{werner}
\een
with
\ben
\rho_{ghz} &=& \vert ghz\rangle\langle ghz\vert,\\
\vert ghz\rangle &=& \frac{1}{\sqrt{2}}]\vert z+ \rangle_1 \otimes \vert z+ \rangle_2\otimes \cdots \vert z+ \rangle_N + \vert z- \rangle_1\otimes \vert z- \rangle_2\otimes \cdots \vert z- \rangle_N ],\\
\rho_{noise} &=& \frac{\mathbb{I}^{\otimes N}}{2^N},
\een
where $ghz$ stands for Greenberger-Horne-Zeilinger \cite{ghz}, and $\mathbb{I}$ is the $2\times 2$ unit matrix. The weight $\gamma$ ($0 < \gamma < 1$) is a free parameter such that the state is entangled if $\gamma > 1/3$ for $N = 2$ and it cannot be described by any local realistic theory if and only if $\gamma > 1/\sqrt{2^{N - 1}}$ \cite{zuk}. Our model is therefore completely specified by the $N$ partite Werner-like state $\rho_W$ \cite{werner} with a single free parameter $\gamma$ and the solutions $A_i (\{{\bf x}_i\},t)$.

Let us consider the special case $N=3$ and the GHZ-like state
\beq
\vert ghz\rangle = \frac{1}{\sqrt{2}}[\vert z+ \rangle_1 \otimes \vert z+ \rangle_2\otimes \vert z+ \rangle_3 - \vert z- \rangle_1 \otimes \vert z- \rangle_2 \otimes \vert z- \rangle_3 ].\label{ghz1}
\eeq
It is an eigenstate of the operator $J_{1x}J_{2x}J_{3x}$ with eigenvalue $-1$. It is also a simultaneous eigenstate of the three hermitian operators $J_{1x}J_{2y}J_{3y}$, $J_{1y}J_{2x}J_{3y}$ and $J_{1y}J_{2y}J_{3x}$ with eigenvalue $+1$ which commute. 
Hence, this state fulfils the following conditions:
\ben
\{J_{1x}\}\{J_{2x}\}\{J_{3x}\} &=& -1,\label{a}\\
\{J_{1x}\}\{J_{2y}\}\{J_{3y}\} &=& 1,\label{b}\\
\{J_{1y}\}\{J_{2x}\}\{J_{3y}\} &=& 1,\label{c}\\
\{J_{1y}\}\{J_{2y}\}\{J_{3x}\} &=& 1. \label{d}
\een
where $\{J_{ix}\} (i=1,2,3)$ denotes the possible outcomes $\pm 1$ of the operation $J_{ix}$ on the single unit $i$ of the state. Notice that every operator occurs exactly twice on the left-hand sides of these four equations, and hence the product of the four left-hand sides must be $+1$. On the other hand, the product of the right-hand sides is $-1$. As explained in Refs. \cite{vaidman} in the context of quantum mechanics, it is possible to assign elements of physical reality ({\em local} and {\em noncontextual} ``hidden variables'') to the underlying variables in such a fashion that they conspire to reproduce the results of the three equations above with $+1$ on the right-hand sides. With such an assignment the prediction for the first equation can only be $+1$ which contradicts the equation which is a prediction for the state. Hence, the state is incompatible with `noncontextual local reality' of the underlying variables. Unlike in quantum mechanics, however, the hidden variables in the brain are only {\em quasi}-hidden, and {\em the incompatibility here implies an emergent property incompatible with the underlying quasi-hidden variable level}.

Now, the amygdala (A), hippocampus (Hip) and the medial prefrontal cortex (mPFC) have been found to play important roles in Pavlovian fear conditioning of rats \cite{maren}. In this form of learning, an emotionally neutral acoustic tone is paired with an aversive stimulus like a footshock. After only a few trials, the rats begin to show a conditioned fear response. If the tone is then repeated many times without the shock, the conditioned fear response decreases. This process is known as {\em extinction}. Pavlov had hypothesized that extinction involves inhibition. Neurophysiological studies have shown that following extinction, neurons of the infralimbic region (IL) of the prefrontal cortex show an increased response to the tone, and this inhibits fear responses in the lateral amygdala (LA) and the central nucleus (Ce) of the amygdala. However, extinction occurs only in the context in which it appeared and not generally. This contextual modulation is effected by the hippocampus either through the LA or the IL. 

Can all this be explained naturally by our model? In order to connect this specific case with our model, we would first have to locate `fear' or equivalently `tenseness' (exemplified by `freezing') within the circumplex plane $XY$ (Fig. 2). It lies close to the positive $Y$ axis (arousal). Hence, the correspondence is clear if we identify LA with the index $i=1$, Hip with the index $i=2$ and IL (of the mPFC) with the index $i=3$. The functions $A_{LA} (\{{\bf x}_i\},t), A_{Hip} (\{{\bf x}_i\},t), A_{IL} (\{{\bf x}_i\},t)$ represent the spatial structures of the three specialized areas of the brain, and their change with time captures their plasticity. Although the state $\rho_B$ is not an eigenstate of the four operators considered above because of the noise term, one can still calculate their expectation values in the state $\rho_W$. It is easy to see that 
\ben 
\langle (J_{LA})_ x(J_ {Hip})_ x(J_ {IL})_ x \rangle &=& - \gamma,\label{A}\\
\langle (J_{LA})_ x(J_ {Hip})_ y(J_ {IL})_ y \rangle &=& \gamma,\label{B}\\
\langle (J_{LA})_ y(J_ {Hip})_ x(J_ {IL})_ y \rangle &=& \gamma,\label{C}\\
\langle (J_{LA})_ y(J_ {Hip})_ y(J_ {IL})_ x \rangle &=& \gamma. \label{D}
\een
The parameter $\gamma$ must then be empirically fitted to see if the tripartite complex is entangled ($\gamma > 1/2$). If it is found to be so under certain conditions, it would undoubtedly signal an emergent coherence that cannot be explained by so-called classical correlations. 

Even before that is done, it is already quite instructive to explore a possible qualitative agreement with the single unit neuroimaging data \cite{maren}. Agreement with the last three equations above can be achieved with the set $\langle(J_{LA})_x\rangle = +$, $\langle(J_{LA})_y\rangle = -$ (no arousal), $\langle(J_{Hip})_x\rangle = -$, $\langle(J_{Hip})_y\rangle = +$ (arousal), $\langle(J_{IL})_x\rangle = -$, $\langle(J_{IL})_y\rangle = +$, but that would contradict the first equation. The first equation would be satisfied if $\langle(J_{Hip})_x\rangle = +$. Thus, the hippocampal valence is a problem for the model unless it is contextual and has a different value in the context of the first equation (only $X$-components, no arousal) from the value it has in the context of the other three equations involving two $Y$-components (arousal). Interestingly, neuroimaging from single unit studies do show that the hippocampus is necessary for context discrimination \cite{phillips}. This contextuality (at the quasi-hidden variable level) can be simply understood as a necessary consequence of nonseparability/entanglement of the LA-Hip-IL complex encoded in the state (\ref{ghz1}). 

Similar analyses can be done with other m-partite ($m < N$) complexes in the brain. The very fact that different complex units of the brain do show remarkable correlations is strongly indicative of their nonseparability/entanglement. Such an understanding of the correlations in terms of a basic Hilbert space structure and collective behaviour of a classical vector field would be {\em complementary} to a phenomenological description of the underlying neurological interconnections among the quasi-hidden constituent units. In other words, the nonseparability/entanglement of the units must be considered as an `emergent phenomenon' that has neural correlates but cannot be wholly reduced to them. 

Before concluding it is necessary to point out a very important difference between classical and quantum measurements and their effect on nonseparability/entanglement. Consider a projection $\Pi \vert \psi\rangle$ where $\vert \psi\rangle$ is a pure state. In quantum mechanics the state $\vert \psi\rangle$ is supposed to `collapse' or `reduce' to the state $\Pi \vert \psi\rangle$ in the sense that the state $(1 - \Pi)\vert \psi\rangle$ disappears! This destroys coherence and entanglement. In classical physics it is not necessary to invoke this hypothesis of `collapse' or `reduction'. All that happens on measurement is that the state $(1 - \Pi)\vert \psi\rangle$ that is not observed or projected is simply blocked or reflected away by the projector such as a polarizer in classical optics. Hence, classical entanglement is robust. There is some empirical evidence that nonseparability/entanglement in the brain is also robust, and is broken only by physical lesions in specialized brain structures. 
 
Quantum measurements may also be described as unitary and reversible operations in a higher dimensional Hilbert space \cite{ghose3}. For example, a state ($\vert +\rangle + \vert -\rangle$) switching spontaneously to either $\vert +\rangle$ or $\vert -\rangle$ can be described by reversible unitary transitions. Any irreversibility that may occur in specific cases can then be ascribed to macroscopic dissipative processes. A brief account of how this is possible is given in the Appendix.

\section{Acknowledgement}
I thank the National Academy of Sciences, India for the award of a Senior Scientist Platinum Jubilee Fellowship which allowed this work to be undertaken. I also thank Sumantra Chattarji for drawing my attention to some literature on Pavlovian fear conditioning in rats.

\section{Appendix}

Let us consider a cbit state 
\beq
\vert \psi\rangle_0 = c_1 \vert 0\rangle + c_2 \vert 1\rangle \label{1}
\eeq with $c_1$ and $c_2$ as membership weights satisfying the condition $\vert c_1 \vert^2 + \vert c_2 \vert^2 = 1$.
Let $M_0$ be a measurement such that 
\beq
M_0 \vert \psi\rangle_0 = \vert 0\rangle
\eeq and $M_1$ a measurement such that 
\beq
M_1 \vert \psi\rangle_0 = \vert 1\rangle.
\eeq These results can also be arrived at by projections which, being many to one, are irreversible. The standard projection operators $\Pi_i (i=0,1)$ are defined by $\Pi_0^2 = \Pi_0, \Pi_1^2 = \Pi_1, \Pi_0 \Pi_1 = \Pi_1\Pi_0 = 0$ and $\Pi_0 + \Pi_1 = \mathbb{I}$. Let us represent the cbit state in the 2-dimensional space spanned by the basis vectors $\vert 0\rangle$ and $\vert 1\rangle$ by the column matrix
\[\vert \psi\rangle_0 =\left(\begin{array}{c}
 c_1\\ c_2
\end{array} \right) \]
The projection operators in this space are reprented by the matrices

\[\Pi_0 = \left(\begin{array}{cc}
 1\,\,0 \\ 0 \,\, 0
\end{array} \right), \,\,\,\,\ \Pi_1 = \left(\begin{array}{cc}
 0\,\,0 \\ 0 \,\, 1
\end{array} \right)\]
which have no inverses, so that
\[\Pi_0 \vert \psi\rangle_0 = c_1 \left(\begin{array}{cc}
1 \\0 
\end{array} \right) = c_1 \vert 0\rangle, \,\,\,\,\ \Pi_1 \vert \psi\rangle_0 = c_2 \left(\begin{array}{cc}
0 \\1
\end{array} \right)= c_2 \vert 1\rangle\]
are irreversible processes. The resulting states have to be normalized by dividing by $c_1$ and $c_2$ respectively for further calculations with them. This is tantamount to redefining the projection operators as $\Pi_0/c_1$ and $\Pi_1/c_2$ which then become state dependent and are no longer idempotent, i.e. $(\Pi_i/c_i)^2 \neq \Pi_i/c_i$.

On the other hand, it is possible to represent the measurements $M_0$ and $M_1$ by the {\em unitary} matrices

\[M_0 = \left(\begin{array}{cc}
 \,\,\,\,c_1^*\,\,\,\,\,\,\, c_2^* \\-c_2 \,\,\,\,\, c_1
\end{array} \right), \,\,\,\,\ M_1 = \left(\begin{array}{cc}
\,\,\,c_2\,\,\,\,\,\,- c_1 \\c_1^* \,\,\,\,\,\,\,\,\,\,\, c_2^* \end{array} \right)\]
so that
\[M_0 \vert \psi\rangle_0 = \left(\begin{array}{cc}
1 \\0 
\end{array} \right) = \vert 0\rangle, \,\,\,\,\ M_1 \vert \psi\rangle_0 = \left(\begin{array}{cc}
0 \\1
\end{array} \right)= \vert 1\rangle.\]
Notice that no further normalization is required as in the case of a projection. The inverses of these matrices are given by 
   
\[M_0^{-1} = M_0^\dagger = \left(\begin{array}{cc}
 \,\,\,c_1\,\,\,\,\,\, - c_2^* \\ c_2 \,\,\,\,\,\,\,\,\,\,\,\,\, c_1^*
\end{array} \right), \,\,\,\,\ M_1^{-1} = M_1^\dagger =  \left(\begin{array}{cc}c_2^*\,\,\,\,\,\,\,\,\,\,c_1 \\- c_1^* \,\,\,\,\,\,\, c_2 \end{array} \right)\]
Notice that $M_0 M_1 \neq 0$, $M_1 M_0 \neq 0$ and $M_0 + M_1 \neq \mathbb{I}$, and hence $M_0$ and $M_1$ are not projection operators.
 
Finally, let us consider the two operators
\beq
U = \left(\begin{array}{cc}
0\,\,\,\,\,\, e^{i\theta} \\ 1\,\,\,\,\,\,\, 0\end{array}\right), \,\,\,\, U^\dagger = \left(\begin{array}{cc}
0\,\,\,\,\,\,\,\, 1 \\ e^{-i\theta}\,\,\,\, 0\end{array}\right).
\eeq Let $c_2 = \vert c_2 \vert e^{i\phi}$. Then, straightforward calculations show that
\beq
U M_0 U^\dagger = M_0^{-1}, \,\,\,\,\,
U M_1 U^\dagger = M_1^{-1}\label{U1}
\eeq provided $\theta = - 2 \phi$.

If one chooses $c_1$ and $c_2$ to be real and equal to $\frac{1}{\sqrt{2}}$, then it is easy to see that 
\ben
M_0 &=& M_1^{-1}= \frac{1}{\sqrt{2}}[\mathbb{I} + i\sigma_y]\label{4}\\
M_1 &=& M_0^{-1} = \frac{1}{\sqrt{2}}[\mathbb{I} - i\sigma_y]\label{5}
\een where $\sigma_y$ is a Pauli matrix, and
\beq
U =  \sigma_x =\left(\begin{array}{cc}
0\,\,\,\,\,\, 1 \\ 1\,\,\,\,\,\,\, 0\end{array}\right).\label{U2}
\eeq

An algebraic geometric way of viewing this is to recognize that it uses the lowest dimensional ($n =2$) Hopf bundle $S^1 \hookrightarrow S^3 \rightarrow S^2$ with fibre $S^1$, base space $S^2$ and total space $S^3$ \cite{sen}. This can be generalized to arbitrary $n$ because quantum mechanics operates on a projective Hilbert space $\mathbb{C}{\rm P}^{n-1}$. The overall phase $U(1)$ ($\sim S^1$) of a state being unphysical can be shrunk to a point. Therefore, the generalized Hopf fibration is $S^1 \hookrightarrow S^{2n-1} \rightarrow \mathbb{C}{\rm P}^{n-1}$. Thus, what looks like projections in $S^{2n-2}$ can be described as unitary transformations on $S^{2n-1}$, and hence reversible.

\end{document}